\documentclass[a4paper,
               ]{jacow}
\usepackage{pdfpages,multirow,ragged2e} %
\usepackage{comment}
\usepackage{fancyhdr}
\usepackage{doi}
\fancypagestyle{firstpage}{
  \fancyhf{} 
  \fancyhead[R]{\small\textsf{FERMILAB-CONF-25-0658-SQMS-TD}}
  
}
\makeatletter%
	\ifboolexpr{bool{xetex}}
	 {\renewcommand{\Gin@extensions}{.pdf,%
	                    .png,.jpg,.bmp,.pict,.tif,.psd,.mac,.sga,.tga,.gif,%
	                    .eps,.ps,%
	                    }}{}
\makeatother

\ifboolexpr{bool{xetex} or bool{luatex}} 
 {}                                      
 {\usepackage[utf8]{inputenc}}           

\usepackage[USenglish]{babel}

\begin{document}
\thispagestyle{firstpage} 
\title{Detection of high-f Gravitational Waves using SRF Cavities}

\author{M. Wenskat\textsuperscript{1,2,}\thanks{marc.wenskat@desy.de}, B. Giaccone\textsuperscript{3,}\thanks{giaccone@fnal.gov}, J. Branlard\textsuperscript{1}, V. Chouhan\textsuperscript{3}, C. Dokuyucu\textsuperscript{1,2},  L. Fischer\textsuperscript{2}, \\ I. Gonin\textsuperscript{3}, A. Grassellino\textsuperscript{3}, W. Hillert\textsuperscript{2}, T. Khabiboulline\textsuperscript{3},  T. Krokotsch\textsuperscript{2}, F. Ludwig\textsuperscript{1}, \\ G. Marconato\textsuperscript{2}, A. Melnychuk\textsuperscript{3}, G. Moortgat-Pick\textsuperscript{1,2}, A. Muhs\textsuperscript{1}, A. Netepenko\textsuperscript{3}, Y. Orlov\textsuperscript{3}, \\ M. Paulsen\textsuperscript{2}, K. Peters\textsuperscript{1}, L. Pfeiffer\textsuperscript{2},
  S. Posen\textsuperscript{3},
O. Pronitchev\textsuperscript{3},
H. Schlarb\textsuperscript{1} \\
\textsuperscript{1}Deutsches Elektronen-Synchrotron DESY, Hamburg, Germany \\
\textsuperscript{2}Universität Hamburg, Hamburg, Germany  \\
\textsuperscript{3}Fermi National Accelerator Laboratory, Batavia, IL, USA
}

\maketitle

\begin{abstract}
Today, apart from some isolated R\&D efforts, there are no gravitational wave (GW) experiments, yet which explore a large part of the vast frequency range above the LIGO/Virgo band. It is planned to establish an experiment at Deutsches Elektronen-Synchrotron (DESY) and at the Superconducting Quantum Materials and Systems (SQMS) Center at Fermi National Accelerator Laboratory (Fermilab) to search for high-frequency GWs in the frequency range of 10\,kHz to 100\,MHz. The basic idea is to use superconducting radiofrequency (SRF) cavities to detect tiny harmonic deformations induced by GWs which change the boundary conditions of the oscillating electromagnetic field.

This paper summarizes the challenging environmental boundary requirements, and the R\&D to operate a cavity using a low level RF (LLRF) system which pushes beyond state-of-the-art accuracy and resolutions and a seismic noise mitigated cryostat at 1.8\,K.

The focus of this paper is the warm and cold commissioning of a prototype cavity, built 20 years ago during the MAGO collaboration, and its first measurement in our collaborative research project.
\end{abstract}

\section{Introduction}
In the search for gravitational waves (GWs) the central focus has been on the Hz to kHz frequency range, which is where the strongest signals from known astrophysical objects were expected. This is the frequency band where the LIGO/Virgo interferometers discovered GWs in 2015 which were produced in the merging of two massive black holes \cite{LIGO2016}. Yet, interferometers are not the only technology developed in the search for GWs.
Electromagnetic (EM) cavities can also be employed in the search for GWs, where the mechanical structure of the cavity itself plays the role of the resonant bar. In this setup, the electromagnetic eigenmodes of the cavity serve as mechanical-to-EM transducers, analogous to Weber bars, where the transducer is an LC circuit.
In this detection concept, an electromagnetic resonator is configured with two nearly degenerate modes, where RF power is injected into only one mode. An incoming GW can transfer power from the loaded mode (0) to the quiet mode ($\pi$) which is maximized when the resonant condition $|\omega_\pi - \omega_0| = \omega_g$ is met. This process, in which signals of two frequencies are combined, is commonly referred to as heterodyne detection. The power transfer is indirectly induced by the deformation of the cavity walls, which leads to the described mode mixing. 

The idea to detect GWs with superconducting radio-frequency (SRF) cavities dates back to the 1970’s. Towards the end of the 70’s Pegoraro et al. \cite{Pegoraro1978a, Pegoraro1978b} and Caves \cite{Caves1979} published papers which proposed the heterodyne detection and the mechanical interaction of GWs with the cavity wall. In the 80’s Reece et al. \cite{Reece1984, Reece1986} started an experimental R\&D programme which was based on the configuration proposed by Pegoraro. This work was based on pillbox cavities and the excited mode was measured through the reflection of the input ports. It was shown that small (order of \SI{e-17}{\centi\metre}) harmonic displacements were detectable with such a superconducting parametric converter. 
 
\begin{figure}[!htbp]
	\centering
		 
  \includegraphics[width=0.8\columnwidth]{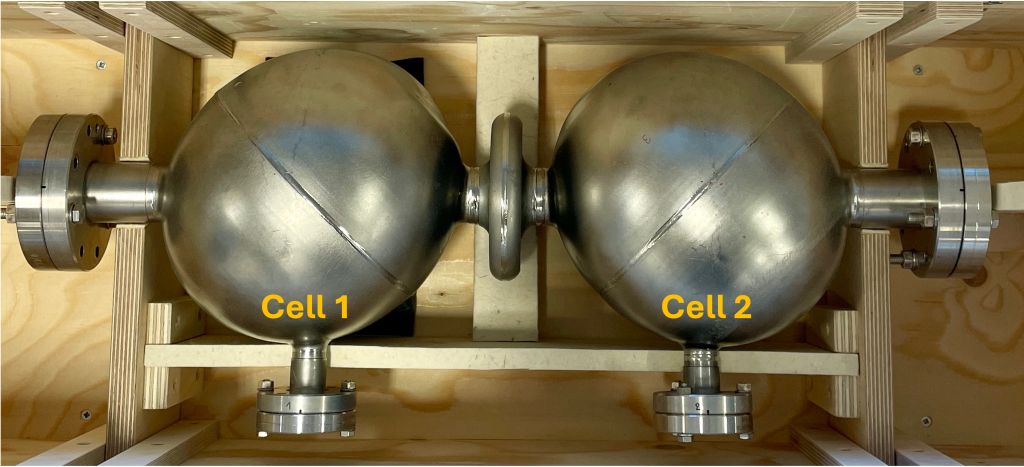}
	\caption{The {\tt PACO-2GHz-variable} prototype niobium spherical cavity. In the picture we highlight which cell we refer as cell 1 vs cell 2.}    	
	\label{fig:MAGO}
\end{figure}

This detection concept was further developed starting at the end of the 90’s within the MAGO proposal with the goal for a scaled-up experiment with 500 MHz cavities as a CERN-INFN collaboration \cite{Ballantini2003, Ballantini2004, Ballantini2005}. Since this proposal stems from the time before the discovery of GWs, the aim was to reach frequencies in the lower kHz range which have sensitivity to astrophysical sources. Although the final project was not funded, three SRF niobium cavities were built during the R\&D activities. The first cavity (a pill-box cavity) was used as a proof-of-principle experiment, which demonstrated the working principle and the development of an RF system to drive and read out the cavity with the necessary precision~\cite{Ballantini2003, Ballantini2004, Ballantini2005}. The second prototype cavity had two spherical cells with fixed coupling. The third cavity, shown in Fig.~\ref{fig:MAGO}, is a spherical 2-cell cavity (denoted {\tt PACO-2GHz-variable}) with an optimized geometry and a tunable coupling cell to change the coupling between the cells and so the frequency difference between the two modes. This cavity was never treated nor tested until now within this project \cite{Fischer2025}, prompted by renewed theoretical interest in this type of setup \cite{Berlin2023}.

\section{Warm Commissioning}
\subsubsection{Geometry}
\begin{figure}[!htbp]
	\centering
		\includegraphics[width=0.95\columnwidth]{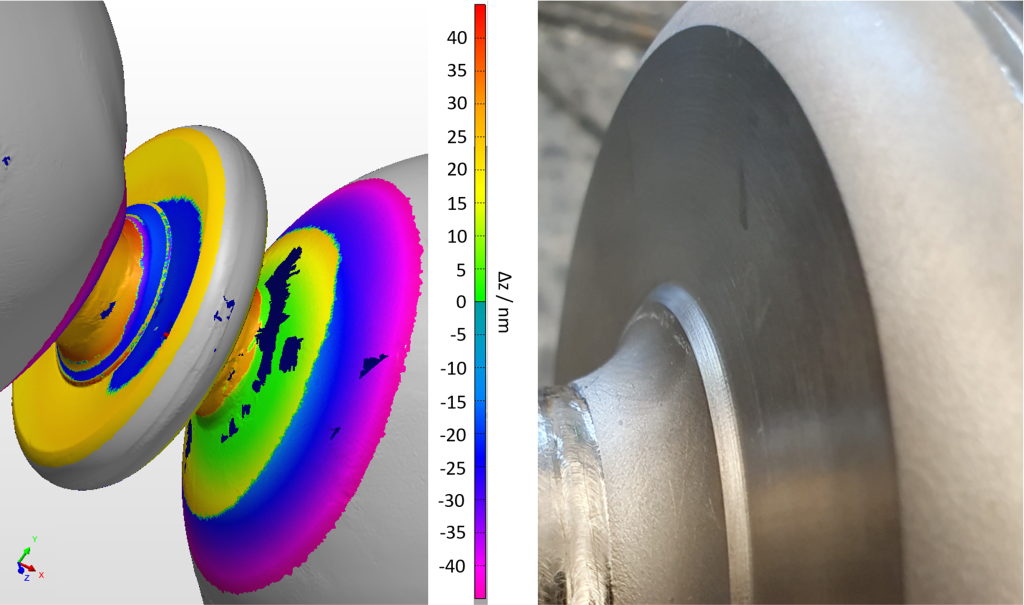}
	\caption{On the left: 3D Scan of the cavity, focused on the coupling cell. The colour code describes the deviation from the nominal cavity geometry. The dent in the tunable cell is clearly visible by the blue area. On the right: Close up image of the coupling cell. The outer surface was turned after deepdrawing to reduce the wall thickness and the force necessary to tune it.}    
	\label{fig:TunableCell}
\end{figure}

\begin{figure}[!htbp]
	\centering
		\includegraphics[width=1.00\columnwidth]{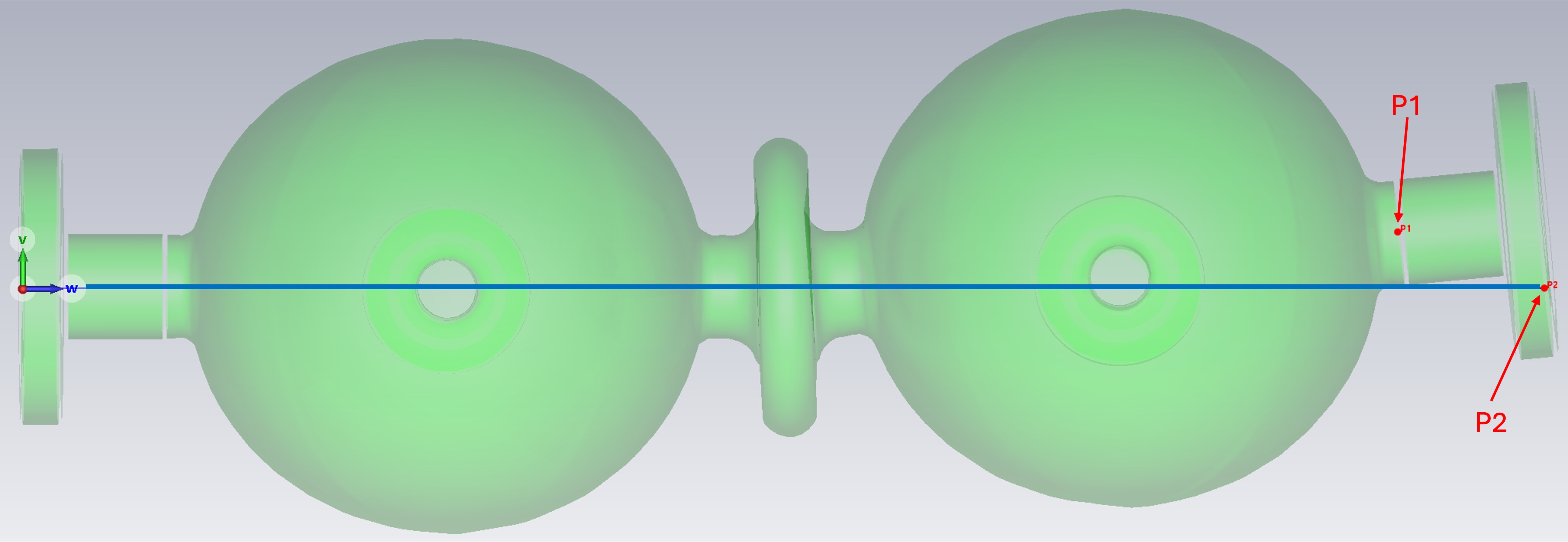}
	\caption{3D visualisation of the cavity. The view is towards the two flanges at the side of each cell. The bending is mostly into one plane and clearly visible. The point $P_1$ on the rotational axis of the tube and the point $P_2$ on the nominal rotational axis of the cavity, highlighted on the right side of the cavity, are nominally expected to sit on the same plane, but are misaligned by 2.3~cm in the vertical plane. For reference, the cavity length, from flange to flange, is approximately 65~cm.}    	
	\label{fig:Shape}
\end{figure}

Since the cavity was fabricated over 20 years ago, limited information was available, and it exhibited significant deviations from the nominal geometry. Therefore, the first crucial step was to conduct a comprehensive survey of both the cavity’s geometry and its wall thickness. The cavity was surveyed using a Hexagon Metrology\texttrademark \, 7-axis portable measuring arm equipped with a RS6-Laserscanner. 
First, in one of the cells, a large toroidal-shaped dent was found around the rotational axis of the prolate spheroid. After optical inspection, it was found that this dent was also seen inside the cavity. Second, a dent on the so-called \textit{coupling cell}, which is the small cell in the center, was found, see Fig.~\ref{fig:TunableCell}.
This dent was aligned with a severe bending of the cavity, starting at the centre, of approximately \ang{6}, shown in Fig.~\ref{fig:Shape}, which is the third strong deviation of the scanned geometry from the nominal design. 
The cavity shape was scanned using the high-speed laser scanner, yielding a 3D polygon model. By surface reconstruction as a part of the reverse-engineering-process, this point cloud was translated into a fully usable STEP file which is the basis for all future simulations, unless it is mentioned that the nominal geometry is used instead. 

Another missing piece of information was the thickness of the cavity walls.  The measurement itself was done using the ultrasonic wall thickness measurement device 38DL PLUS from the company Evident Europe\texttrademark\:using the probe V260-SM with a test frequency 10 MHz. 
For the two spherical cells of the MAGO cavity, a highly non-uniform wall thickness was found. The average wall thickness for cell 1 was found to be $(1.84\pm0.05)$\,mm and for cell 2 to be $(1.89\pm0.05)$\,mm. Point to point variations on the cells of \SI{100}{\micro\metre} was observed, and the welding seams were thicker, as expected. The wall thickness of all the tubes were found to be $(1.91\pm0.02)$\,mm.  
The coupling cell was dedicated as such -- a mechanically tunable cell to squeeze the cavity longitudinally, and hence bring the two RF cells closer to each other. This would increase the overlap of the RF fields within the cells, and hence increase the cell-to-cell coupling $k_{cc}$. To minimise the force needed to tune the cavity, and to assure that the deformation takes place at the coupling cell and not elsewhere on the cavity, the wall thickness was intentionally reduced. The wall thickness measurement just before the step (shown in Fig.~\ref{fig:TunableCell}) showed an average wall thickness of $(1.0\pm0.1)$\,mm. This is in agreement with the observed wall thicknesses of the RF cells corrected by the estimated removal of the turning. 
\subsubsection{RF Fields}
\begin{figure}
     \centering
     \includegraphics[width=1.0\columnwidth]{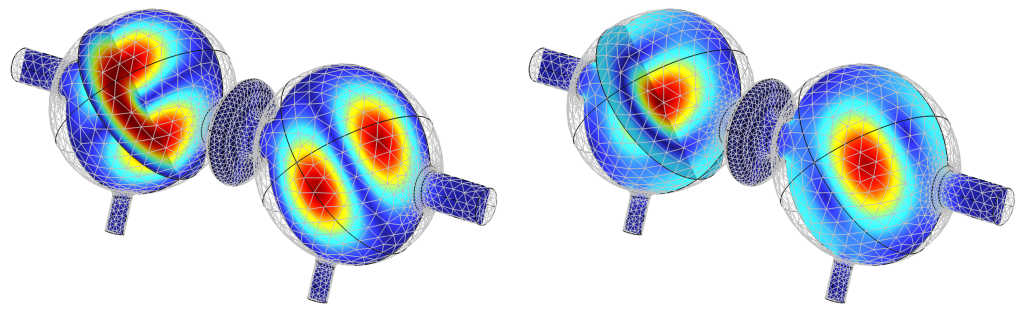}
     \caption{Electric (left) and magnetic (right) field norm of the symmetric and anti-symmetric $\text{TE}_{011}$ modes which only differ in phase ($0$ and $\pi$) and have identical field patterns.}
     \label{fig:RFields}
 \end{figure}

\begin{figure}
     \centering
     \includegraphics[width=1.0\columnwidth]{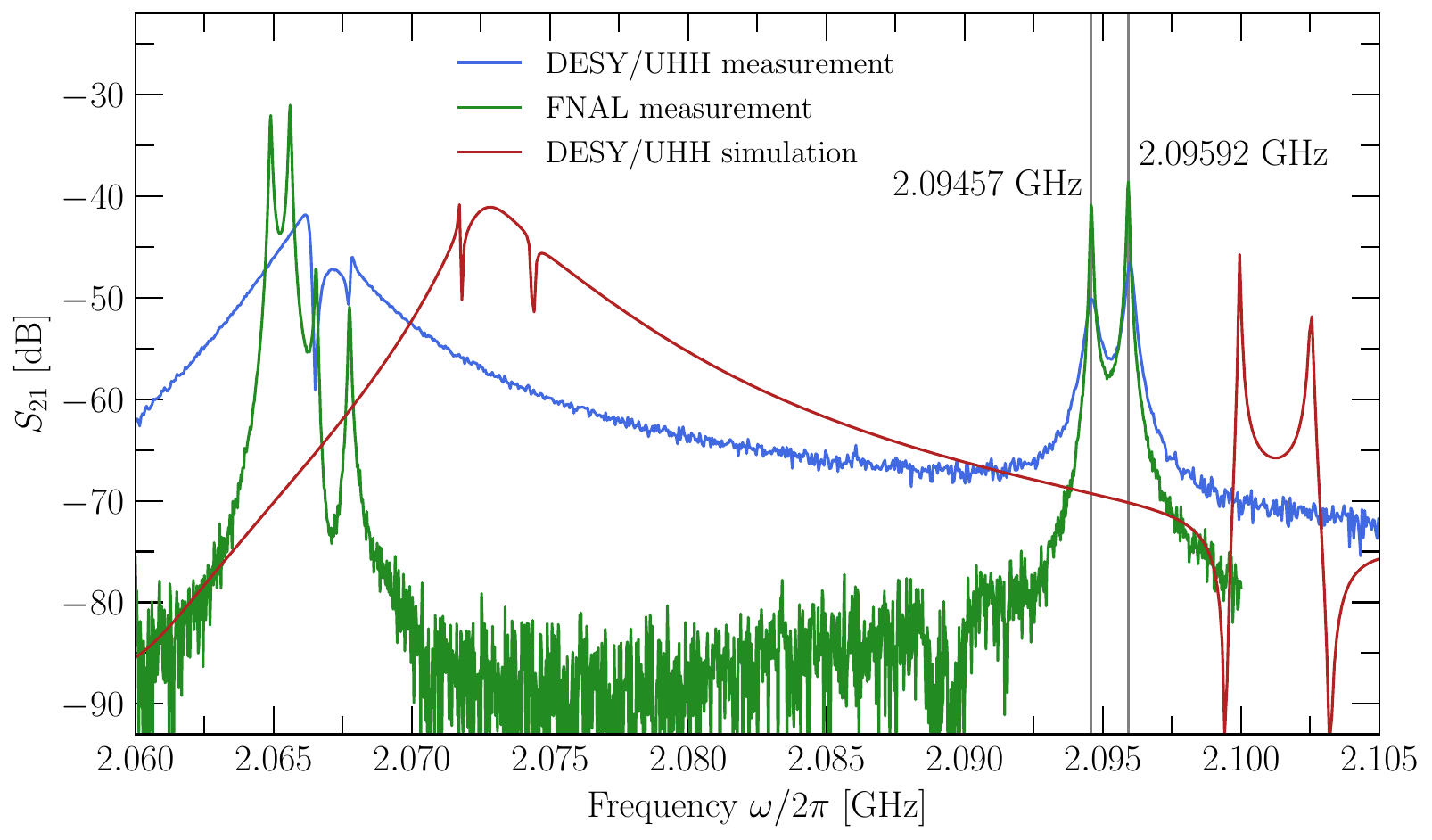}
     \caption{$S_{21}$ measurements and simulation of the cavity at room temperature. Measurements at DESY/UHH (blue) have been performed with pin-antennas which couples to the electric field of the modes, while the measurement at Fermilab (green) used loop-antennas which couples to the magnetic field of the modes. The simulations (red) have been performed using the DESY antenna setup, using the scanned geometry and an averaged wall thickness.}
     \label{fig:VNA_initialspectra}
 \end{figure}

The $\text{TE}_{011}$ mode is the electromagnetic mode of interest for the GWs search with the MAGO prototype cavity. When coupled together, the two nominally identical spherical cells give rise to a family of three pairs of $\text{TE}_{011}$ quasi-degenerate symmetric and anti-symmetric modes. For this search, we are interested in the third pair of $\text{TE}_{011}$ (field profiles are shown in Fig.~\ref{fig:RFields}), and we will refer to the two modes with $0$ or $\pi$ phase difference as symmetric and anti-symmetric respectively. The frequency splitting between the symmetric and anti-symmetric modes can be controlled by tuning the coupling cell. Our simulations based on the nominal cavity geometry yield a resonance frequency of $\omega\simeq2\pi\times\SI{2.1}{GHz}$ for the $\text{TE}_{011}$ modes of interest, and the expected splitting between the symmetric and anti-symmetric modes of $\Delta\omega\sim 2\pi\times\SI{10}{kHz}$. A simulation of the eigenmode spectrum for the scanned geometry revealed that the eigenfrequencies of the $\text{TE}_{011}$ mode in the two cells are separated by more than $\SI{1}{\mega\hertz}$, as plotted in Fig.~\ref{fig:VNA_initialspectra}, and have a poor field flatness with an amplitude ratio on the order of $\mathcal{O}(10^2)$. The geometry of the two spherical cells deviate significantly, causing their respective eigenfrequencies to be too far apart for the weak cell-to-cell coupling. As a result, the symmetric and anti-symmetric modes are primarily determined by the individual cell frequencies.

This was also confirmed with first RF measurements at room temperature, and theoretically explained based on RLC equivalent circuit models. 
We can estimate the cell-to-cell coupling $k_{cc}$ from the measured eigenfrequencies of the coupled system $\omega_0$ and $\omega_\pi$ using the relation
\begin{equation}
    k_{cc}=\frac{\omega_\pi-\omega_0}{(\omega_\pi+\omega_0)/2}\,.
    \label{eq:kcc}
\end{equation}
 However, if the coupling strength is too low for the two single cells having different parameters, the eigenfrequencies of the coupled system converge to the eigenfrequencies of the single cavity cells. This also means that the value of $k_{cc}$ can no longer be inferred from $\omega_0$ and $\omega_\pi$ alone. As both the spectrum of the single cavity cells and of the coupled system were measured, we concluded that the MAGO cavity was actually in that regime and that the two cells were not coupling well and the measured frequencies were the eigenfrequencies of the individual cells. 
 A low $k_{cc}$ is inherent to the detection scheme, as the frequency different between the eigenmode of the coupled cavities should be on the order of 10 kHz, which gives the range for the detection of gravitational waves, while the operational frequency is 2.1\,GHz for now, but in the hundreds of MHz later. Therefore, a coupling of $k_{cc}\leq 10^{-4}$ is expected, which puts tight requirements of the shape accuracy and the similarity of the eigenfrequencies of the cells. Fitting the measured data from Fig. \ref{fig:VNA_initialspectra} yielded eigenfrequencies of the two cells of ${\omega}_1=2.09457$\,GHz and ${\omega}_2=2.09597$\,GHz, showing a difference in the eigenfrequencies of 1.4\,MHz, and a cell-to-cell coupling $k_{cc}$ of $2\cdot 10^{-4}$. Overall, the study of the equivalent circuit leads us to two conclusions. First, for such weakly coupled cavity cells to act as a coupled system, there are stringent requirements on how similar both cavity cells need to be manufactured. Second, the MAGO prototype cavity as we received it did not meet these requirements and needed to be tuned in order to restore the gravitational wave sensitivity it was designed to have.

\section{Surface Treatment and\\ Plastic Tuning} 
In order to improve the merging of the two cells into a coupled system, for the given cell-to-cell coupling $k_{cc}$, it was necessary to plastically deform the individual cells and therefore tune their RF eigenfrequencies. This would then decrease the observed difference between the two eigenfrequencies from the order of $\mathcal{O}(\text{MHz})$ down to the intended $\mathcal{O}(\text{kHz})$.

First, the bend in the cavity, shown in Fig.~\ref{fig:Shape}, was eliminated in order to facilitate preparation steps of the inner cavity surface such as surface chemistry and high pressure rinsing in preparation for the cold RF test.
Due to the unique cavity shape with narrow ``beamtube" apertures, narrow opening for the coupling cell, and wide almost-spherical cells, it was not possible to use electropolishing (EP) to remove material from the inner surface, as EP would require the insertion of a custom cathode in the RF volume. For this reason we planned to rely on buffered chemical polishing (BCP) for the material removal. Given the shape of the cavity, we expected the material removal to be non-uniform and to be higher in areas such as the coupling cell, where the material thickness was already reduced to \SI{1}{\milli\metre}. Additionally, BCP is usually conducted with a rotational system that allows to rotate the cavity while the acid is flown through to avoid gas accumulation on the niobium surface, which may lead to surface defects and increased roughness. However, given the unique cavity geometry the existing facilities didn't allow to carry out a rotational BCP. As a result, it was decided to conduct only a flash BCP with the goal of removing few micrometers (4--\SI{6}{\micro\metre}). The BCP was conducted at the SRF facilities at Argonne National Laboratory. Afterwards, the cavity underwent heat treatment both to remove hydrogen that may have been introduced during fabrication and to relieve internal stresses. The cavity flanges are made of stainless steel and are brazed on the niobium tubes. As a result, the temperature of the heat treatment was limited to 600\,$^{\circ}$C for 24 hours to avoid contamination to the cavity and the furnace. 
\begin{figure}[!htbp]
	\centering
		\includegraphics[width=0.85\columnwidth]{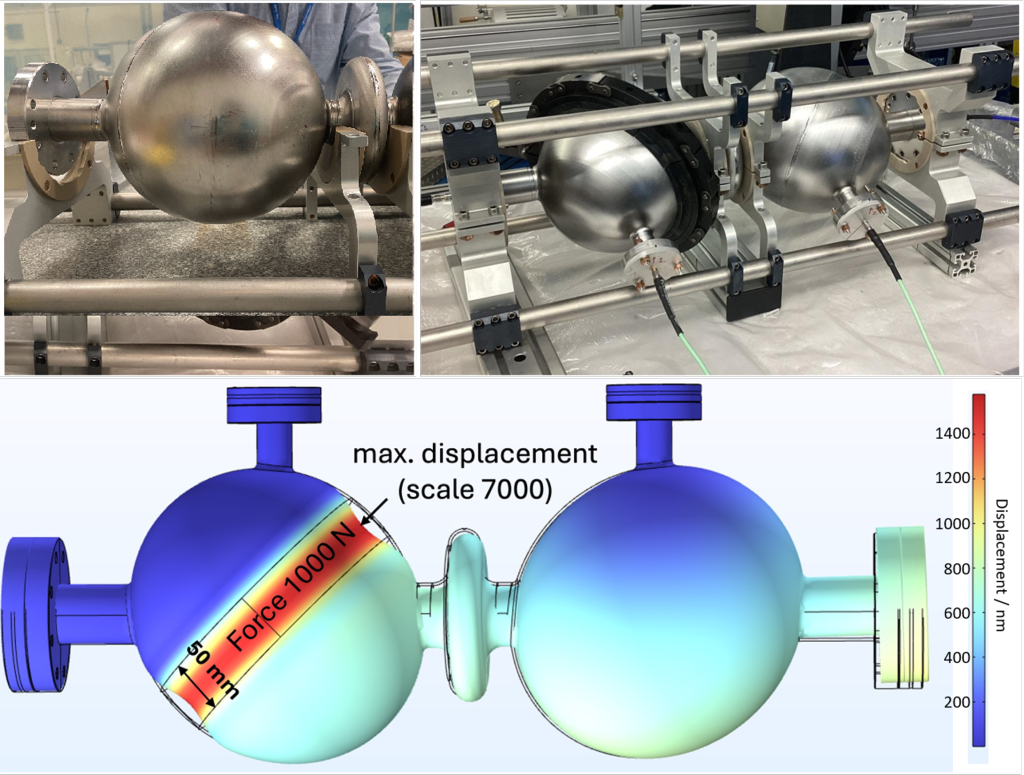}
	\caption{Top left shows the cavity as installed on the bottom half of the frame, prior to straightening: the left flange does not rest on the frame due to the cavity bending. The top right shows the chain tuner installed around the minor axis of cell 1.  Bottom: Deformation study on the RF cell, simulating the effect of the chain-tuner. The strategy was to squeeze a waist around the minor axis of the ellipsoid.}
	\label{fig:TuningSim}
\end{figure}

At Fermilab, the cavity was constrained in a frame designed to support the cavity during the RF cold tests. Removing one of the external support spiders, it was possible to delicately push on the cavity flange until the bent was removed and the cavity was plastically straightened, see top left of Fig.~\ref{fig:TuningSim}. Comparative measurements of the $S_{ij}$ parameters of the cavity before and after each step showed no measurable change in the eigenfrequencies. The cavity was assembled with four small loop antennas for this step, and the following tuning, to monitor the frequency changes via a 4-port vector network analyzer.

\begin{figure}[!htbp]
	\centering
		\includegraphics[width=0.9\columnwidth]{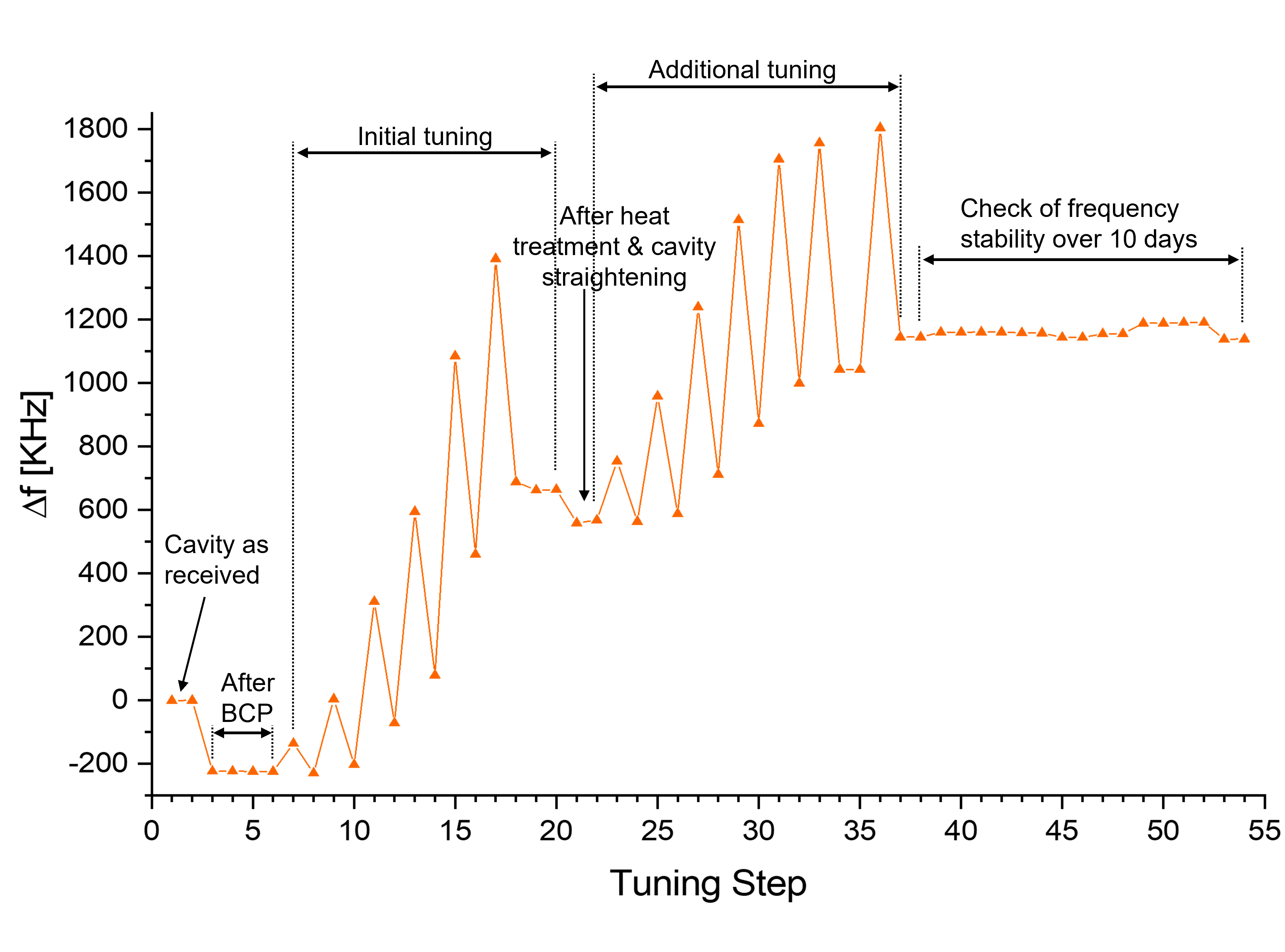}
	\caption{Frequency change vs. tuning steps for the relevant TE\textsubscript{011} eigenmodes. The alternating tightening and relaxing approach used during the tuning procedure is clearly visible. A final increase of +1.2\,MHz is achieved.}    	
	\label{fig:Tuning}
\end{figure}

To tune cell 1, the cell with the lower eigenfrequency, we plastically deformed it by squeezing the minor axis, since the TE\textsubscript{011} mode is the most sensitive to shape deviations in this region, see Fig.~\ref{fig:TuningSim}.
Figure~\ref{fig:Tuning} shows the cell 1 frequency progression starting from the arrival of the cavity at Fermilab up to the completion of the plastic tuning of the two cells. A frequency shift by 200\,kHz was introduced by the inner surface removal of $(4-6)\,\mu\text{m}$, which results in a sensitivity of 40\,kHz/$\mu$m, which is in the same order of magnitude of the sensitivity value (20\,kHz/$\mu$m) obtained from the simulations. Cell 1 was then tuned in multiple steps and an increase of the eigenfrequency by +1.2\,MHz was achieved. As this was more than intended, cell 2 was then also tuned by +200\,kHz. The subsequent measurements (steps $>$ 44) were carried out to assess the stability of the cell tuning over 10 days. Small frequency variations are due to day to day temperature changes in the RF laboratory.
After tuning, the two peaks merge due to the low quality factor at room temperature, which causes the resonances to be broad and overlap. Measuring the frequency of each cell individually, as uncoupled system, shows that the frequency difference at room temperature is now on the order of (4-7)\,kHz.

\section{2\,K RF Cold Test at Fermilab}
\begin{figure}[!htbp]
	\centering
		\includegraphics[width=1\columnwidth]{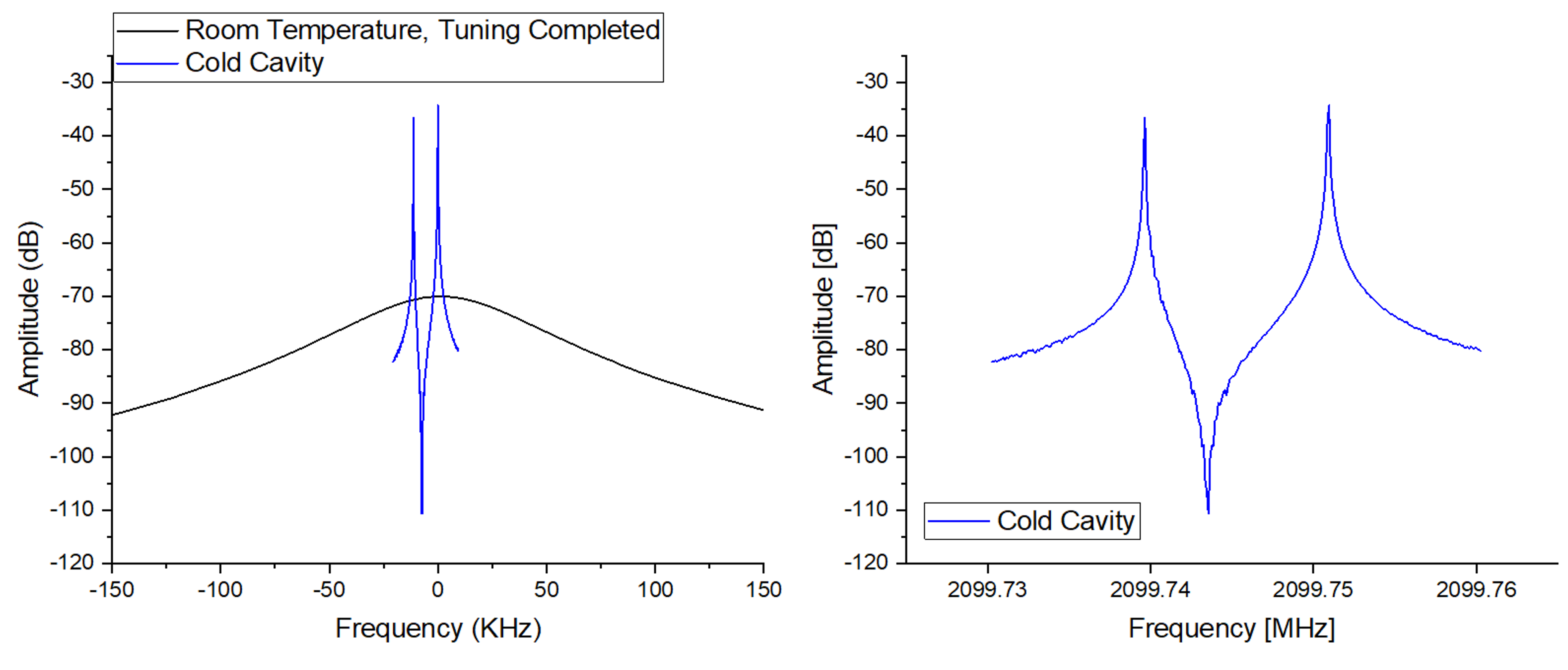}
	\caption{$S_{21}$ measurements of the cavity at room temperature and at 4\,K. With the superconducting transition, the broad peak splits into two high-Q peaks corresponding to the $0$ and $\pi$ modes. The two modes are about 11\,kHz apart, as expected.}    	
	\label{fig:VNA_WarmCold}
\end{figure}

In preparation for the first RF cold test, the cavity was assembled with three loop antennas to couple to the $\text{TE}_{011}$ magnetic mode. One critically coupled antenna (with measured $\text{Q}_{ext1} = 1.6\times 10^{10}$) was installed on the ``beamtube" axis, and the field probes (with measured $\text{Q}_{ext 2,3} = 2-3\times 10^{12}$) were installed in the cell flanges on the transversal plane. The cavity underwent high pressure rinsing and evacuation, and was then installed on the Fermilab vertical insert, equipped with 8 temperature sensors, 5 flux gates, and Helmholtz coils for zero field cooldown. The cavity frequency was monitored during cooldown. As shown in Fig.~\ref{fig:VNA_WarmCold}, below the superconducting transition, as the Q increased, the broad $\text{TE}_{011}$ peak split into two distinct peaks, separated by approximately 11\,kHz, in good agreement with the designed value.

\begin{figure}[!htbp]
	\centering
		\includegraphics[width=1\columnwidth]{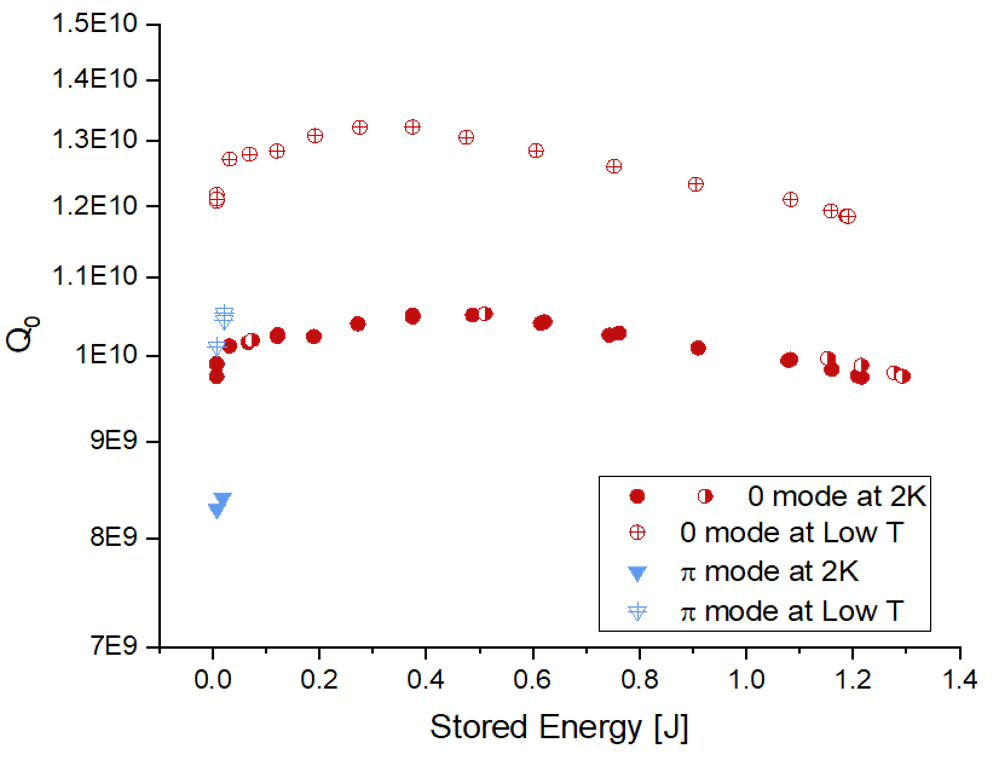}
	\caption{Quality factor vs. stored energy for the two eigenmodes - achieving the intended quality factor with this prototype, almost identical to its predecessor cavity.}    	
	\label{fig:QvE}
\end{figure}

The symmetric and antisymmetric modes (or $0$ and $\pi$ respectively) were measured at 2\,K and at $\approx$ 1.4\,K. Figure\,\ref{fig:QvE} summarizes the performance of the two modes. The cavity exhibits high Q for both modes, in line with the performance of the previous MAGO prototype cavity\,\cite{Ballantini2005}. The symmetric mode quenched at 1.3 J (hard quench), while the antisymmetric mode could only be measured at low energy due to the phase locked loop (PLL) instability. Because the two modes are close in frequency, both fall within the PLL bandwidth. When locked on the $\pi$ mode, increasing input power caused the $0$ peak to emerge above background and the $\pi$ mode to decay, as the PLL preferentially locked to the $0$ mode. Follow-up studies showed this behavior was specific to the configuration used: the input and transmitted antennas were both on cell 1, while the pickup on cell 2 was connected to an HOM port. In this setup, the PLL sees the $0$ and $\pi$ modes with the same phase and locks to whichever mode the pickup couples to more strongly. This issue should be resolved by using the pickup on cell 2 as the transmitted power input for the PLL. In the course of this test, we characterized additional noise sources, including RF input phase noise, amplifier noise, and microphonics; these results are not presented here.

\section{4\,K RF Cold Test at DESY}
\begin{figure}[!htbp]
	\centering
		\includegraphics[width=1\columnwidth]{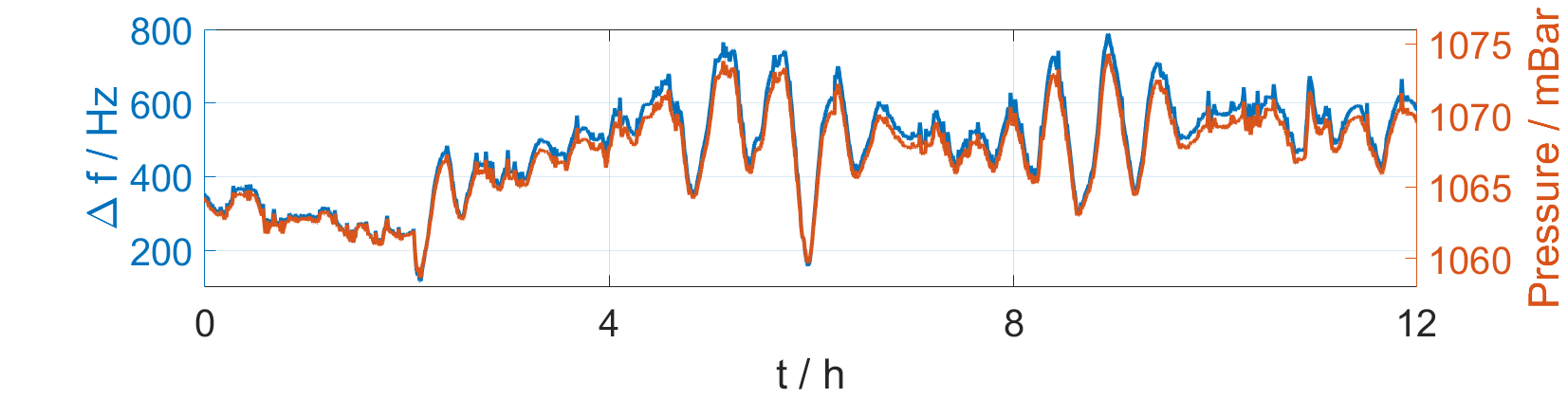}
	\caption{Pressure (orange) and frequency detuning (blue) vs. time for the 0-mode over a period of 12\,h at 4\,K during which the cryo-plant had significant stability issues. This clearly shows that the LLRF system can track the cavity and keep the system on resonance over a significant bandwidth. From this, a $df/dp$ for the 0-mode of 42.8\,Hz/mbar was obtained.}    	
	\label{fig:Drift}
\end{figure}

\begin{figure}[!htbp]
	\centering
		\includegraphics[width=1\columnwidth]{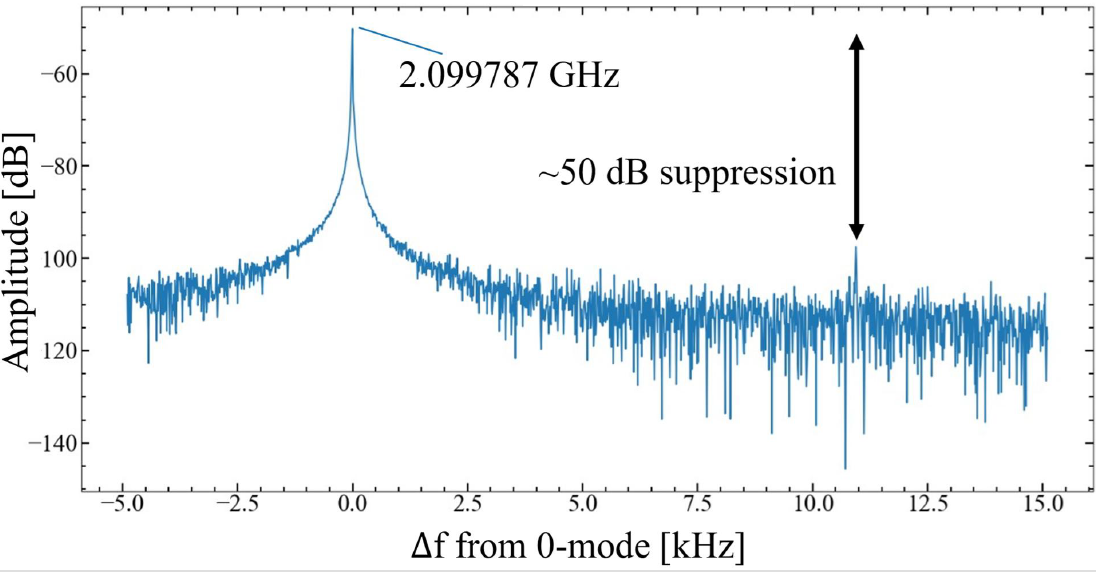}
	\caption{Test of the mode rejection concept: spectrum of the pick-up signal by driving both cells simultaneously with a split-up signal, but one signal was phase-shifted to minimize the $\pi$-mode excitation. With this simple make-shift set-up, a suppression by 50\,dB in a single-stage setup was achieved -  compared to the shown 140\,dB of a two-stage setup of the PACO collaboration}    	
	\label{fig:Suppression}
\end{figure}

After the successful test at Fermilab, the cavity was shipped fully assembled and under vacuum to DESY. After an incoming inspection, the cavity was installed in an insert, and an extensive 4\,K test campaign for the low-level RF (LLRF) development was carried out in preparation for the 2\,K cold test. 
The aim for the 4\,K test campaign was to test the newly developed $\upmu$TCA system for the 2.1\,GHz frequency and to check various RF properties of the cavity - antennas system, and only two major results are presented here.
Figure \ref{fig:Drift} shows the frequency and pressure versus time for a period of 13\,h during which the cavity was driven with the $\upmu$TCA. The pressure variations during that time caused a pressure drift of nearly 18\,mbar peak-to-peak. This caused a frequency shift of the 0-mode of up to 700\,Hz peak-to-peak. From this, two results can be obtained. First, a pressure sensitivity of 42.8\,Hz/mbar was obtained and, second and more important, that the LLRF system was able to track the frequency changes and kept the cavity on resonance the whole time through these severe detuning. This shows the successful implementation of the cavity control algorithm which was able to keep the cavity on resonance the whole time.
Another objective was to test the suppression of the signal - or $\pi$- mode with a simple, make-shift set up. As the cavity has two pick-up antennas - one for each cell - and one input antenna, we installed a phase-shifter into the branch of the pick-up signal from the cell which was not driven, and shifted the phase of that signal. This way, a suppression of the $\pi$-mode excitation by 50\,dB in a single-stage setup was achieved (see Fig.~\ref{fig:Suppression}), compared to the shown 140\,dB of a two-stage setup of the PACO collaboration \cite{Ballantini2005}.

\section{CONCLUSION AND OUTLOOK}
In this paper we presented the first results of the revival of this SRF cavity based R$\&$D program for high frequency gravitational waves. We successfully characterized the cavity at room temperature, prepared it for the first cold test and carried out RF cold tests at 2\,K by SQMS at Fermilab and at 4\,K at DESY. The next immediate steps are a 2\,K test at DESY to further develop the LLRF system and include a new developed 2.1\,GHz Carrier-Suppression Interferometer (CSI) to improve the sensitivity by further suppressing the noise floor \cite{Springer2022}. Also, piezo actuators and sensors will be attached to the cavity to mimic a gravitational wave signal and therefore test the detection scheme, but also to measure the mechanical resonances of the cavity at 2\,K as this is a crucial information for the overall detection scheme \cite{Marconato2025}. On an intermediate term, a better suspended insert and a different antenna configuration will be deployed, and a physics run for first exclusion limits in a yet unexplored area is planned. Based on these studies, a new cavity with an optimized geometry will be fabricated, after which another physics run will be carried out.

\section{ACKNOWLEDGEMENTS}
For the loan of the {\tt PACO-2GHz-variable} cavity, we thank the Istituto Nazionale di Fisica Nucleare, Italy.
The authors thank Asher Berlin, Sergio Calatroni, Andrea Chincarini, Sebastian Ellis, Gianluca Gemme, Roni Harnik, Cornelius Martens, Andreas Ringwald, Udai Raj Singh, Jan Hendrick Thie and Hans Weise for their support and useful discussions. 
This material is based upon work supported by the U.S. Department of Energy, Office of Science, National Quantum Information Science Research Centers, Superconducting Quantum Materials and Systems Center (SQMS) under contract number DE-AC02-07CH11359. This manuscript has been authored by FermiForward Discovery Group, LLC under Contract No. 89243024CSC000002 with the U.S. Department of Energy, Office of Science, Office of High Energy Physics. LF, WH, TK, GM-P, KP and MW acknowledge support by the BMBF under the research grant 05H21GURB2 and the project is also funded/acknowledges support by the Deutsche Forschungsgemeinschaft (DFG, German Research Foundation) under Germanys Excellence Strategy - EXC 2121 `Quantum Universe' - 390833306.

\end{document}